\documentclass[pra,twocolumn,superscriptaddress,amsmath,amssymb]{revtex4}

\usepackage{url}
\usepackage{subfigure}

\usepackage{graphicx,amsthm,amsfonts}
\newcommand{\be}{\begin{equation}}
\newcommand{\ee}{\end{equation}}
\newcommand{\beq}{\begin{eqnarray}}
\newcommand{\eeq}{\end{eqnarray}}

\begin{document}

\title{Lattice gas simulations of dynamical geometry in two dimensions}
\author{Anna Klales}
\author{Donato Cianci} 
\author{Zachary Needell}
\affiliation{Department of Physics, 370 Lancaster Ave., Haverford College, Haverford, PA 19041 USA.}
\author{David A. Meyer} 
\affiliation{Department of Mathematics, UCSD, La Jolla, CA 92093 USA}
\author{Peter J. Love}
\affiliation{Department of Physics, 370 Lancaster Ave., Haverford College, Haverford, PA  19041 USA.}
\affiliation{Institute for Quantum Information, California Institute of Technology, Pasadena, CA 91125 USA}

\begin{abstract}
We present a hydrodynamic lattice gas model for two-dimensional flows on curved surfaces with dynamical geometry.  This model is an extension to two dimensions of the dynamical geometry lattice gas model previously studied in one-dimension~\cite{1D1,1D2,1D3}.  We expand upon a variation of the two-dimensional flat space FHP model created by Frisch, Hasslacher and Pomeau, and independently by Wolfram~\cite{FHP,wolfram}, and modified by Boghosian, Love, and Meyer in~\cite{peter}.  We define a hydrodynamic lattice gas model on an arbitrary triangulation, whose flat space limit is the FHP model. Rules that change the geometry are constructed using the Pachner moves, which alter the triangulation but not the topology~\cite{Pachner}.  We present results on the growth of the number of triangles as a function of time. Simulations show that the number of triangles  lattice grows with time as $t^\frac{1}{3}$, in agreement a mean field prediction. We also present preliminary results on the distribution of curvature over a typical triangulation for these simulations.
\end{abstract}

\maketitle

\section{Introduction}

Lattice-gas automata (LGA) models for fluids date from the sixties, when Kadanoff and Swift and Hardy, de Pazzis and Pomeau introduced the first such models~\cite{KS,HPP}. Both of these models use a two-dimensional Cartesian lattice and are anisotropic. Since simple fluids are isotropic, these models are not capable of reproducing hydrodynamics.  This problem was solved in 1986 when Frish, Hasslacher, and Pomeau, and independently Wolfram, introduced an isotropic model (the FHP model) using a triangular lattice.  They demonstrated that an LGA models the Navier-Stokes equations in flat two-dimensional space~\cite{FHP,wolfram}.

All LGA methods are characterized by phases of propagation and collision of particles that move on a lattice.  During the propagation phase, particles move from site to site on the lattice, while during the collision phase the particles rearrange themselves amongst the vectors at each site (see Figure~\ref{notation}).  Before we discuss the FHP rules in detail, it is important to note that the rules that govern these models are not meant to replicate the physical world on a small scale;  the Navier-Stokes equations emerge from the FHP rules on the macroscopic scale for large lattice sizes and spatial or ensemble averaging.  The microscopic rules are only required to conserve total momentum, particle number, and energy.  Additionally, the lattice must be sufficiently symmetric to yield an isotropic pressure tensor.

Many 2-D situations of physical interest use a Euclidean plane as the underlying geometry, hence ``lattice'' gases, in which the model is constructed on a translation invariant discretization of Euclidean space. However situations exist, such as atmospheric flow, the experiments of Seychelles~\cite{seychelles:144501}, or surface flows in interfaces embedded in fluid mixtures, in which a discretization of a sphere or other surface in which the geometry is non-Euclidean may be more appropriate. In such geometries, the angles of a triangle need not sum to $\pi$. We may specialize to simplicial complexes made up of equilateral triangles, as any 2-D surface may be discretized in this way~\cite{Pachner}. In this case the geometry is defined locally by the number of triangles meeting at each grid point. If six triangles meet, the geometry is locally flat. If fewer than six triangles meet, the geometry has positive local curvature. If more than six triangles meet, the geometry has negative local curvature. If the properties of the triangulation, including the local curvature, are allowed to change we call the geometry dynamical.

There are many situations in physics in which geometry takes on a dynamical role. Perhaps the most fundamental is in Einstein's general theory of relativity, in which the idea of motion along geodesics in a Riemannian manifold supervenes Newtonian ideas of acceleration due to forces~\cite{GR}. In the Regge treatment of general relativity~\cite{Regge} and the causal dynamical triangulations approach to quantum gravity~\cite{Ambjorn,Loll} these Riemannian manifolds are replaced by simplicial complexes. The statistical mechanics and growth dynamics of random surfaces has been much studied for both 1-D interfaces~\cite{KPZ} and 2-D surfaces~\cite{KKN1,KKN2,Wheater, PACZUSKI}. In spite of their origin in very different physical systems, the common language of discretized surfaces can be informative. For example, the crumpling transition of membranes~\cite{PACZUSKI} also occurs in Euclidean approaches to simplical quantum gravity~\cite{Ambjorn}.

In this paper we present a hydrodynamic lattice gas model for two-dimensional flows on curved surfaces with dynamical geometry.  We extend a variation of the FHP model to arbitrary equilateral triangulations. We allow the geometry so defined to become dynamical by applying the Pachner moves contingent on the particle content.  The restriction of time-reversibility is used to restrict the rule space, as in the one dimensional version of this model~\cite{1D1,1D2,1D3}. We present a mean-field prediction and simulation results for the growth of the lattice as a function of time, and give preliminary results on the distribution of curvature on the triangulations generated by these simulations. We close the paper with some conclusions and directions for future work. 

\section{The FHP Model}

\begin{figure}[htp]
\includegraphics[width = 3.2 in]{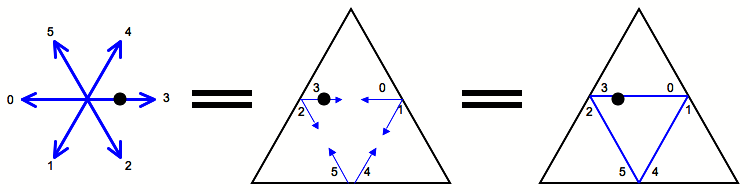}
\caption{An FHP lattice site has 6 possible velocities labeled 0 through 5, each of which represents the velocity of a particle.  Each vector can hold at most one particle, so that each site has $2^6 = 64$ states.  The traditional representation of a site in the FHP model is the star shown on the left.  By moving the vectors to the edges of a triangle, as shown in the center picture, we convert the site from a single point to the face of a triangle.  These two sites are equivalent.  If we remove the arrow heads from the vectors, we produce the notationally convenient right hand figure.  We refer to the conversion between the star and the triangle as inflation.}
\label{notation}
\end{figure}

In the FHP lattice gas automata the particles move on a triangular lattice. At each lattice site there are six lattice vectors. Each vector can be occupied by at most one particle - the model has an exclusion principle. The vector which a particle occupies defines its velocity. In Figure~\ref{notation}, for example, the site is occupied by a single particle moving to the right. The state of a particular site is given by its particle content.  Each vector at each site can have two states: occupied or unoccupied. This gives a total of $2^6 = 64$ states per site.  

To facilitate the generalization from a Euclidean lattice to an arbitrary triangulation, we would like to regard our sites as triangles rather than a single point.  We therefore enclose the site in a triangle and move the vectors to the edges of the triangles, as demonstrated by Figure~\ref{notation}.  This ``inflated'' site is equivalent to the point site used by Frisch, Hasslacher and Pomeau. This modification of the FHP model was originally proposed by Boghosian, Love, and Meyer~\cite{peter}.  Those authors proposed collisions on the edges of the triangles, where four vectors meet and introduced the possibility of having a rest particle of mass two in the model for a total of five bits per state. This model was analyzed by a grouping of triplets of triangle edges sites followed by a Chapman-Enskog expansion which yielded isotropic fluid equations.

\begin{figure}[htp]
\includegraphics[width = 2.0 in]{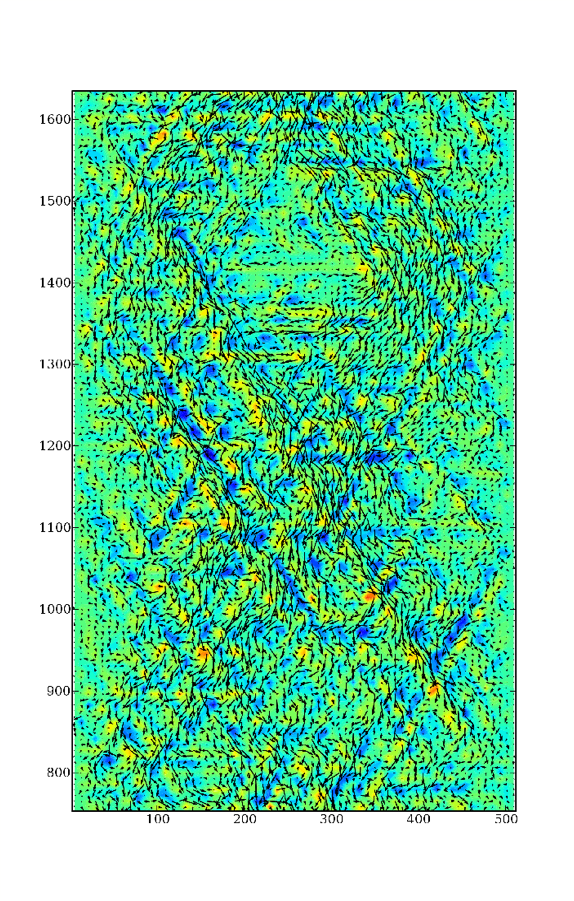}
\caption{Channel flow simulation with a barrier using the model defined in~\cite{peter}. Arrows indicate the velocity field, and the color scale indicates vorticity. In this formulation of the model collisions occur on the edges, rather than the faces, of the triangulation. This leads to separate conservation of momentum along each lattice direction and hence an anisotropic model. This anisotropy is evident in this simulation as the structure of the lattice is visible in the flow.\label{bad}}
\end{figure}

However, running a channel flow simulation using their proposed model produced the image in Figure~\ref{bad}.  Note that the structure of the lattice is evident on a macroscopic scale in the figure.  This is due to a spurious conserved momentum in collisions at the vertices of the Kagome lattice. No momentum is transferred between separate lines of the lattice, so momentum is conserved in three directions in two-dimensional space.  This leads to unphysical flows, an example of which is shown in Figure~\ref{bad}.  We therefore redefine the FHP model with collisions occurring on the faces, rather than the edges, of the triangle. It should be noted that the Chapman-Enskog analysis presented in~\cite{peter} remains valid for a model, such as the one we present here, in which momentum is exchanged by collisions among all lattice directions. 

\begin{figure}[htp]
\includegraphics[width = 3.0in]{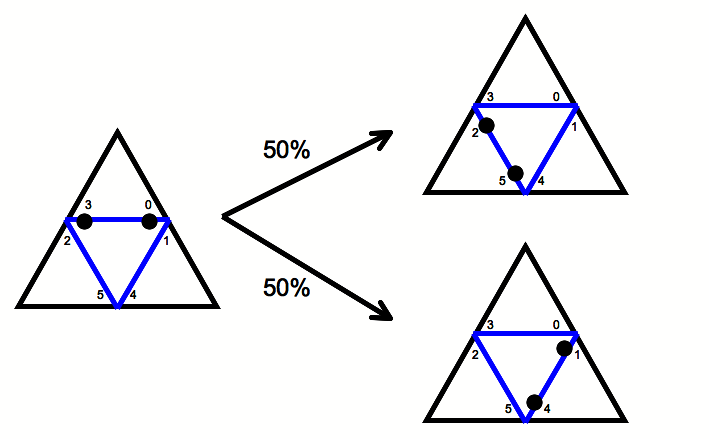}
\caption{A two-particle collision at a site.  The particles switch with equal probability to one of the other two directions of the lattice.  This rule applies to any two particles entering a site with opposite velocities.}
\label{2body}
\end{figure}

\begin{figure}[htp]
\includegraphics[width = 3.0in]{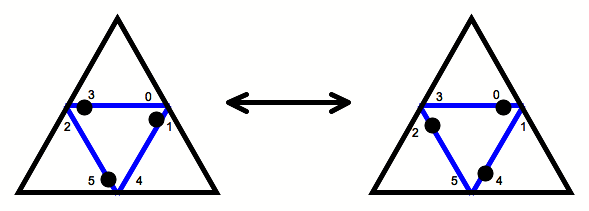}
\caption{A three-particle collision at a site.  The particles switch to the unoccupied vectors of the site.  This move breaks the separate conservation of momentum along each direction of the lattice.}
\label{3body}
\end{figure}

The rules used for fixed geometry in the variant of the FHP model we study are shown graphically in Figures~\ref{2body} and~\ref{3body}. If two particles enter a site with opposite velocities, as in Figure~\ref{2body}, they flip to either of the other lines of the lattice with equal probability. If three particles enter a site such that their total momentum sums to zero, in other words, there is a particle occupying every other vector, the particles switch from the occupied vectors to the unoccupied vectors, shown in Figure~\ref{3body}. This three-body collision breaks the separate conservation of momentum along each line of the lattice. This is required because extra conserved quantities lead to incorrect macroscopic behavior. If particles enter in any other configuration, they are simply allowed to propagate as usual to the next site along their geodesic~\cite{FHP}.

\section{Lattice gases on curved surfaces}

\begin{figure}
\includegraphics[width = 3.0 in]{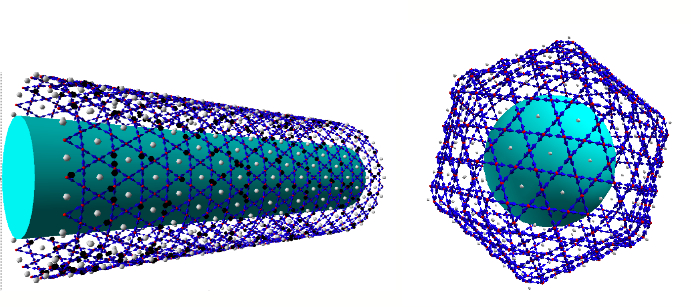}
\caption{A Kagome lattice on a  cylinder (left) and on an icosahedron (right).  The nodes of the triangulation are depicted as white spheres.  Any triangulation of a two-dimensional manifold can be tiled with a Kagome lattice.  Particles are allowed to move along the lines of the Kagome lattice, and are shown in black on the left. These images were generated with visual python~\cite{snake}.}
\label{kagomeduo}
\end{figure}

We now generalize the FHP model to arbitrary equilateral triangulations. It is known that any manifold can be approximated arbitrarily closely by a tiling of equilateral triangles~\cite{Pachner}. This allows us to triangulate any surface and regard each face as an inflated FHP site. In the special case of flat space the vectors in the array of inflated sites (Figure~\ref{notation}) create a tiling of Stars of David, along the lines of which the particles can move.  This lattice is known as the Kagome lattice.  Figure~\ref{kagomeduo} shows a triangulation of a cylinder and an icosahedron where the nodes of the triangulation are shown in white and the nodes of the Kagome lattice are in red. These images were generated with visual python~\cite{snake}. We now describe the rules which couple the particles to the triangulation and allow the geometry to become dynamical.

To allow the geometry to become dynamical, we employ the Pachner moves. A sequence of Pachner moves cannot change the topology of a manifold, but it can take the manifold from one triangulation to another: a torus can morph into another toroidal geometry such as a coffee mug, but it can not morph into a sphere~\cite{Pachner}. The state of the system with static and flat geometry is specified by the particle content of the sites alone, the state of the system with dynamical geometry specified by both the particle content and the geometry of the triangulation. 

\begin{figure}
\includegraphics[width = 3.0 in]{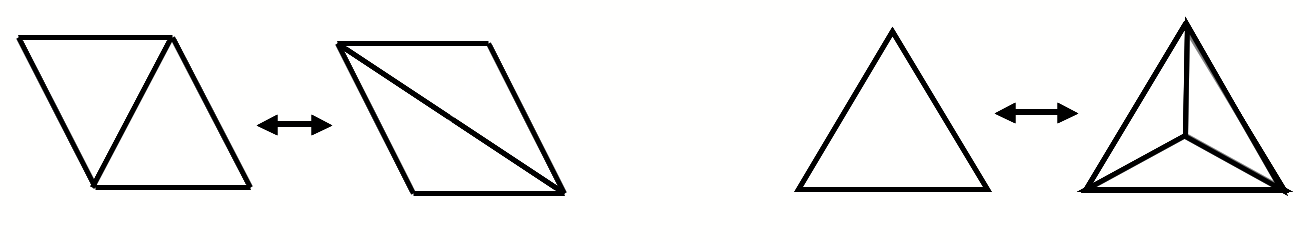}
\caption{Pachner Moves.  A sequence of Pachner moves can connect any pair of triangulations of a manifold, but cannot change the topology.  The two-to-two move (left) changes the orientation of two triangles as show above, and the one-to-three move (right) replaces one triangle with three, creating a tetrahedron, or {\em vice versa}.}
\label{Pachner}
\end{figure}

There are two Pachner moves for two dimensional triangulations: a two-to-two move, where the number of triangles is unchanged, and a one-to-three or three-to-one move that increases or decreases the number of triangles by two. We call the three-to-one and the one-to-three move addition and deletion, respectively, because they add or subtract a tetrahedron from the surface. 

In three dimensions the two-to-two Pachner move (Figure~\ref{Pachner}, left) is not isometrically embeddable in general. If two triangles are removed, turned, and replaced in the triangulation, they will not fit unless the dihedral angle between the original pair of triangles was that of the tetrahedron.  This is unsurprising since two-dimensional manifolds are not generically embeddable in three dimensions~\cite{nash}. However, the one-to-three move, (Figure~\ref{Pachner}, right), is generically immersible, although it is not generically embeddable because it may cause self-intersection of the surface.

\begin{figure}[h]
\includegraphics[width = 3.0in]{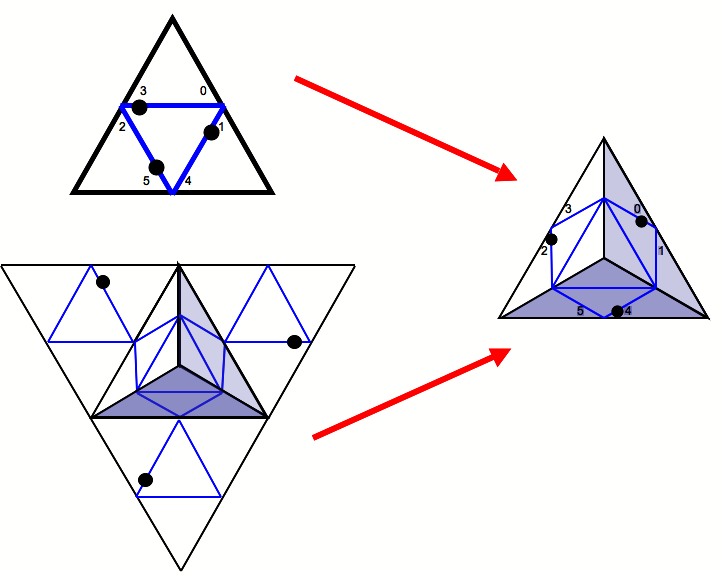}
\caption{New geometry must be created empty so that one state does not have two preimages, a problem illustrated in this figure.  Both figures on the left would produce the figure on the right if they were allowed to time evolve.  The particles on the triangle in the upper left hand corner are therefore required to propagate off before the new geometry is created.  This allows rules with dynamical geometry in which every state has a unique preimage, and which are therefore invertible.}
\label{timereversal}
\end{figure}

To couple the flow to the geometry we must specify how the application of a particular Pachner move is triggered by the particle content. The rules for fixed geometry involve particles on a single triangle. The locality of a rule which changes the geometry is determined by the locality of the Pachner moves. The one-to-three, three-to-one and two-to-two moves are triggered by the state of one, three or two triangles respectively. As in one dimension, the constraint of time reversibility is applied in order to restrict the set of rules considered~\cite{1D1}. We first recall the distinction between invertibility and reversibility. A rule is invertible if every state has a unique preimage ---  given the state (particle content plus geometry) one may reconstruct the whole unique history leading to that state for an invertible rule. For a reversible lattice-gas rule, the history of a given state may be generated by an inverse rule which can be interpreted as propagation and the same collision rules that generate the forward time evolution. One must recall that the inverse of the product of collision and propagation $CP$, is $P^{-1}C^{-1}$.

We choose to apply a one-to-three Pachner move after a three particle collision. The particles undergo the three-body collision of Figure~\ref{3body}, and then propagate to neighboring triangles. The Pachner move is applied in the subsequent collision. The restriction of time-reversibility is satisfied if we create new geometry after the particles have propagated from the triangle. That is, we do not create new geometry that contains particles. If we created new geometry with particle content, the resulting state may have two preimages: one preimage in which the geometry is about to be created, and one preimage in which particles are about to advect onto existing geometry. This problem is illustrated in Figure~\ref{timereversal}.  

\begin{figure}[h]
\includegraphics[width = 3.0in]{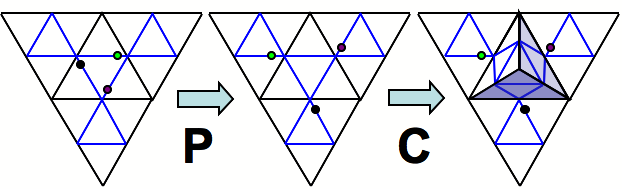}
\caption{Creation. A triangle is replaced by a tetrahedron. This collision is triggered by a three particle collision, where three particles enter a site on every other vector (all odd numbered vectors or all even numbered vectors) such that the combined momentum of the three particles is zero.  The particles then propagate away, and the tetrahedron is formed.}
\label{Creation}
\end{figure}

\begin{figure}[h]
\includegraphics[width = 3.0in]{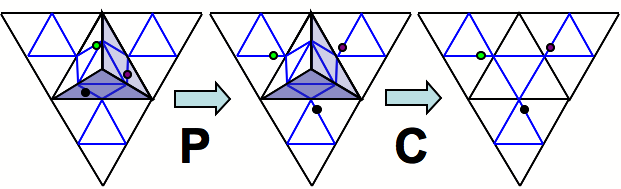}
\caption{Deletion. Three triangles forming a tetrahedron are replaced by a single triangle.  This collision is triggered when three particles propagate off the tetrahedron, as shown on the left above.  This also happens when the particles leave the tetrahedron on the empty vectors that point to the surrounding triangles.}
\label{Deletion}
\end{figure}

The rules for addition and deletion are illustrated in Figures~\ref{Creation} and~\ref{Deletion}.  When three particles enter a triangle on all even numbered vectors or all odd numbered vectors and then leave, that triangle is replaced by a tetrahedron (see Figure~\ref{Creation}).  When three particles propagate off a tetrahedron in the same manner, the tetrahedron is deleted (see Figure~\ref{Deletion}).

\begin{center}
 \begin{figure}[h]
 \includegraphics[width = 3.0 in]{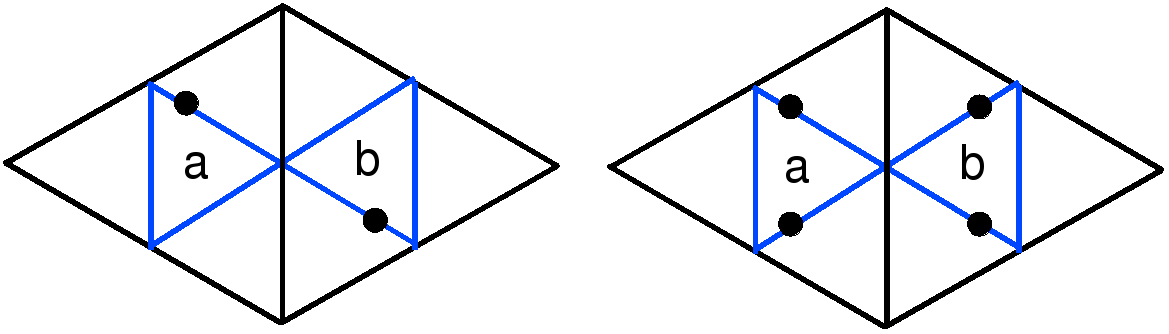}
\caption{The two to two move is triggered by two different states:  the two particle state, left, and the four particle state, right. The particles remain where they are during these moves.}
\label{trigger}
\end{figure}
\end{center}

We now determine rules for applying the two-to-two move. This move does not create new geometry and so it is straightforward to ensure that the rule is time-reversible.  This move is triggered by two different states: a four-particle state, and a two-particle state, shown in Figure~\ref{trigger}.

The rules including dynamical geometry are therefore modified from the stochastic FHP rules defined in Figures~\ref{2body} and~\ref{3body} by the fact that the three body rule of Figure~\ref{3body} is followed by the one-to-three Pachner move shown in Figure~\ref{Creation}. The stochastic two-body rules remain unchanged. The rule set also includes the two-to-two Pachner moves shown in Figure~\ref{trigger} in which the geometry changes by the particle states do not. Naturally, the inclusion of the stochastic two body rule of Figure~\ref{2body} renders the model as a whole irreversible. This could be remedied by the addition of a rest particle and the replacement of the stochastic two body rule by a deterministic rule as described in~\cite{peter}. Here we avoid the use of a rest particle and retain the stochastic two-body rule.

The virtue of requiring reversibility of the dynamical geometry rules is that the resulting model is fundamental, allowing study of the origin of thermodynamics in its classical version, and in principal allowing a natural generalization to a quantum version. Reversibility also constrains the rule space to allow definition of a simple and relatively natural model. 

\section{Implementation}

Implementation of the rules defined in the previous section presents several challenges. In this section, we describe some of the details of our implementation which allow the model to be efficiently simulated without reference to embedding space coordinates. 

We first distinguish extrinsic geometry from intrinsic geometry.  When specifying a triangulation, one can use an extrinsic definition of the geometry, or an intrinsic definition. An extrinsic definition describes the triangulation by relating it to an ambient or embedding space. For example, a tetrahedron can be defined extrinsically by giving the Cartesian coordinates of its vertices in three dimensional Euclidean space.  

Geometry can also be defined intrinsically, without reference to embedding in some higher dimensional space.  For example, we can define a tetrahedron intrinsically as follows.  First, we specify that a triangle is defined by three points equidistant from each other.  Then, we specify that we have four triangles, and that each triangle shares exactly one edge with every other triangle. This is illustrated in Figure~\ref{squid}.  We have defined a tetrahedron intrinsically.  There was no reference to any coordinate system, only reference to parts of the triangulation itself. 

For a lattice gas model defined on an arbitrary triangulation the flux of particles defines a velocity field. A velocity vector on a triangulation lives in the tangent space to the triangulation at that point. In general, transport of tangent vectors on manifolds requires a description of the relationship between tangent spaces at different points on the manifold. For example, when computing a covariant derivative on a Riemannian manifold one must consider the variation of coordinate basis vectors with position on the manifold. The components of the derivatives with respect to the coordinates of the basis vectors are the Christoffel symbols, which specify the connection on the manifold. These quantities are intrinsic: they may be computed from the metric without reference to any higher dimensional embedding space.

The implementation of our model contains both intrinsic and extrinsic geometry information. The extrinsic (embedding space) information is the set of vertex coordinates, velocity vectors, and particle coordinates of our two dimensional simulation in three dimensional space.  This is used to produce visualizations of the model, such as those shown in Figures~\ref{bad} and~\ref{kagomeduo}. It is possible to imagine situations in which the extrinsic information is coupled to the intrinsic model dynamics. For example, a membrane embedded in a bulk fluid will have dynamics driven in part by the embedding space fluid dynamics. We only consider model dynamics defined intrinsically. In particular this means that the dynamics remains perfectly well defined even if the triangulation is no longer isometrically embedable in three dimensional Euclidean space. We do include the possibility that moves which would be allowed by the intrinsic dynamics are forbidden conditioned on the embedding, however, for all simulations described in this paper these constraints were inactive.

The collision rules are defined locally and must conserve mass, momentum and energy of the particles. We wish to apply the same collision rule on every triangle expressed in terms of the vector labels. In general, translation of a triangle from one location on the triangulation to another will induce a transformation of the vector labels. A reflection of the vector labels through one of the symmetry axes of the triangle will change the definition of momentum between one triangle and another. Because of this we restrict to labelings in which the transformation relating the vector labellings of any two triangles is one of the three proper rotations of the labeled triangle shown in Figure~\ref{notation}. 

The propagation rule moves particles from one triangle to another. This operation depends on the transformation of the labeling of vectors on going from one triangle to another. For each triangle each of the six vectors carries two pieces of connectivity information which define this transformation. Firstly each vector carries a triangle label which gives the triangle reached by propagation along that vector. Secondly each vector carries a vector label which determines the vector the particle arrives at after propagation. Because the vector labellings of any two sites are related by one of three rotations, the labelling of two adjacent triangles is determined by the image of any one of the vectors. Hence the connectivity information redundantly determines the geometry.

\subsection{Implementing the two-to-two Pachner move}

\begin{figure}[htp]
 \includegraphics[width = 3.0 in]{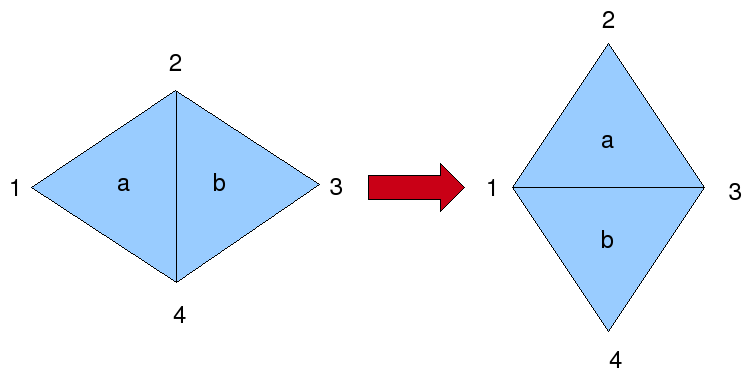}
\caption{The two-to-two Pachner move.  The pair of triangles is rotated in the lattice, so that vertex $1$ goes to $2$, vertex $2$ goes to $3$, vertex $3$ goes to $4$, and vertex $4$ goes to $1$. The relationship between triangles ${\bf a}$ and ${\bf b}$ stays the same; only the connectivity between each triangle and the surrounding triangles is redefined, along with the vertices of each triangle.\label{2to2}
}
\end{figure}

The two-to-two Pachner move changes the orientation of two neighboring triangles but not the number of triangles. After a collision applying such a move it is necessary to change the connectivity information of the surrounding triangles. The move is shown in Figure~\ref{2to2}, where triangles {\bf a} and {\bf b} form a rhombus.  During the two-to-two move, the four vertices of the rhombus undergo a cyclic permutation as the rhombus rotates. The connectivity between the two triangles involved in the move and the surrounding triangles must be updated, and the positions of the vertices in embedding space will change unless the dihedral angle between ${\bf a}$ and ${\bf b}$ is that of the icosahedron.  The  connectivity between triangles {\bf a} and {\bf b} does not change. 

Triangle pairs which are candidates for the two-to-two move are identified before the propagation phase.  Triangles in the appropriate states, for example the state of triangle {\bf a} in Figure~\ref{trigger}, are identified. Then, their partner triangle is examined to see if it in the the state shown in triangle {\bf b} in Figure~\ref{trigger}.  If it does, the move is performed; connectivity is redefined with the surrounding triangles and the vertices of the triangles are updated.  

\begin{figure}[htp]
 \includegraphics[width = 2.0 in]{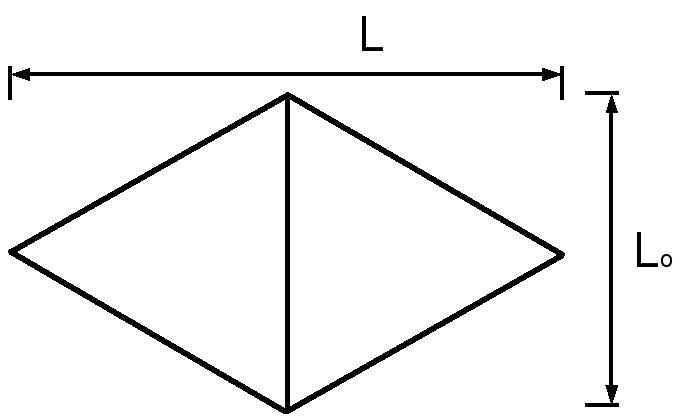}
\caption{If the new edge, $L$, is more than a fraction $x$ different from $L_0$, the two-to-two move is prevented.\label{embed}}
\end{figure}

As noted above, unless the dihedral angle of the two original triangles is that of the tetrahedron, applying the two-to-two move will result in a triangulation which is not embeddable in three dimensional Euclidean space. A control exists in the code which prevents the change in the embedded length of edges from deviating from the equilateral value by more than a specified fraction. If $L_0$ is significantly different from $L$ as shown in Figure~\ref{embed}, i.e., the new edge is significantly different from the new edge, we do not allow the two-to-two move to be performed.  Given a fraction $x$, $0< x\leq 1$, we determine whether or not the change will be performed via the restriction
\begin{equation}
\left(1- x\right)L_0 < |L| < \left(1+x\right)L_0.
\end{equation}

What values of $x$ are relevant? For an initially flat triangulation, $L=\sqrt{3}L_0$, and for a single tetrahedron added to an initially flat triangulation $L=\sqrt{2}L_0$. For triangles which meet with the dihedral angle of the icosahedron $L=\phi L_0/2$ where $\phi$ is the golden ratio. Hence for $x< (\sqrt{2}-1)$ with an initially flat triangulation no two-to-two moves will be performed. For $x< (\sqrt{3}-1)$ positive curvature added to an initially flat triangulation by a one-to-three move is frozen in place at the new vertex, as no two-to-two moves may be applied involving any face of the tetrahedron. For $x< (\phi/2-1)$ no two-to-two moves may be applied to an icosahedron. 

The effect of these moves is therefore to increase the edge lengths of the triangles according the the Euclidean metric in three-dimensional embedding space. Either one may imagine the triangulation inflated by a scale factor and embedded isometrically in a higher dimensional space in such a way that the three dimensional embedding is a projection of this higher dimensional embedding. In this case the restriction specified by $x$ is to triangulations whose projections into three dimensional Euclidean space are almost isometric. Alternatively, one may regard the three dimensional embedding space itself as no longer Euclidean. In this case, $x$ represents a bound on the deviation of the metric of the three-dimensional embedding space from Euclidean. Note that because $x$ specifies a ratio between new and old embedded lengths this constraint allows triangle edge lengths grow repeatedly by a succession of geometry changing moves. One could also implement a constraint which would bound all embedded edge lengths above by an additive constant.

\subsection{Implementing the one-to-three Pachner move: Addition.}

\begin{figure}[htp]
\includegraphics[width = 3.0 in]{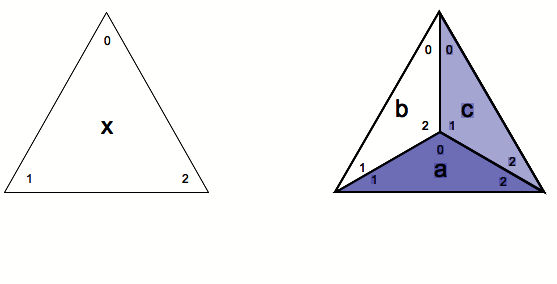}
\caption{The $1-3$ Pachner move (creation). The triangles are labelled ${\bf x}$, ${\bf a}$, ${\bf b}$, and ${\bf c}$, and the vertices by $0$, $1$ and $2$.  (${\bf m}_n$ refers to triangle $m$'s $nth$ vertex).  When triangle ${\bf x}$ is triggered for addition, copies of triangle ${\bf x}$ are rotated out of the plane of the paper along each edge.  For example, to produce triangle ${\bf b}$ triangle ${\bf x}$ is copied and rotated along the edge joined by the vertices ${\bf x}_1$ and ${\bf x}_0$, matching vertex ${\bf x}_2$ with the apex. Acting similarly for the other two sides, triangle ${\bf x}$ on the left transitions to the tetrahedron on the right.  Deletion is the inverse of this process, and intersections of common vertices and triangle neighbors are used to identify the relevant vertices. The apex is identified as the intersection of the vertices of triangles ${\bf a}$, ${\bf b}$, and ${\bf c}$, since the apex is the only vertex shared by all three triangles. \label{tetra1-3}}
\end{figure}

Triangles triggered for addition are marked before propagation, and undergo changes in geometry during collision. First, the embedding space coordinates of the apex of the new tetrahedron are determined. The three triangles of the new tetrahedron each have two of the vertices of the triggered triangle and the third vertex is the apex. The move may be regarded as making three copies of the original triangle and ``rotating'' each triangle along a different edge so that its free vertex becomes the apex.  In Figure~\ref{tetra1-3}, triangle {\bf a} has been rotated along the $1-2$ edge, triangle {\bf b} has been rotated along the $0-1$ edge, and triangle {\bf c} has been rotated along the $0-2$ edge.  One of the three new triangles replaces the original. The connectivity of the new tetrahedron is set to be that shown in Figure~\ref{deletionex} where ${\bf a}$ is the  original triangle and ${\bf b}$ and ${\bf c}$ are the two added triangles.

The curvature at any triangle vertex is equal to six minus the number of triangles meeting at that vertex. If the one-to-three Pachner move is implemented without restriction, vertices of the triangulation with arbitrarily large negative curvature may form. This is because the one-to-three move adds a new vertex with positive curvature and increases the number of triangles meeting at each of the original three vertices of the triangle by one. In order to allow simulations in which the curvature is bounded between $\pm c$ we forbid addition of tetrahedra on a triangle with any vertex with curvature $c$. Bounding the curvature to be $\pm 1$ from the original triangulation is equivalent to preventing new tetrahedra from forming on existing tetrahedra.  

\subsection{Implementing the 3-1 Pachner move: Deletion.}

\begin{figure}[htp]
\includegraphics[width = 3.0 in]{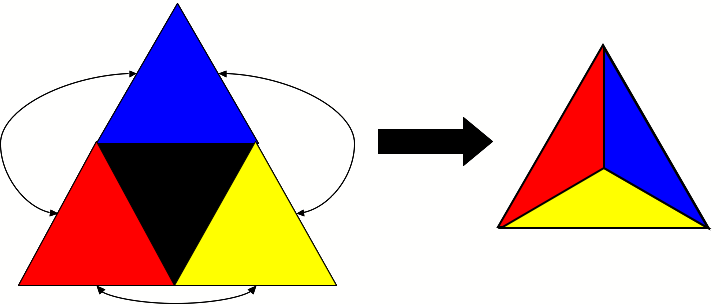}
\caption{A triangle whose neighbors are neighbors belongs to a tetrahedron.  This is an intrinsic definition of a tetrahedron.  On the left there are three triangles sharing one edge each with the black triangle.  They are the black triangle's neighbors. If the blue (top), red (bottom left), and yellow (bottom right) triangles in the left figure are also neighbors as designated by the arrows above, the three triangles fold into a tetrahedron, right.\label{squid}
}
\end{figure}

The deletion rule depends on the state of three triangles in a tetrahedral configuration. Tetrahedra are identified as sets of triangles whose neighbors are neighbors using the intrinsic information - this test uniquely specifies a tetrahedron (see Figure~\ref{squid}). Once tetrahedra have been identified, they are checked to see if their particle content makes them a pre-image of deletion.  

Deletion presents a computational issue, in that triangles must be removed from the list in which they are stored. This would change the indices of all triangles, requiring a relabeling of the whole triangulation, a computationally demanding process. It is more efficient to disregard the triangles that have been deleted, and place them on a dummy list. These obsolete triangles are ignored whenever the state is updated, and hence relabeling is avoided.  To prevent the list of triangles from expanding too quickly due to addition and deletion, one of the three triangles in a deleted tetrahedron is replaced with the single replacement triangle, placing the other two on the dummy list. This is the inverse of the  one-to-three addition move in which the original single triangle becomes one of the triangles of the new tetrahedron. 

\begin{center}
\begin{figure}[htp]
\includegraphics[width = 3.0 in]{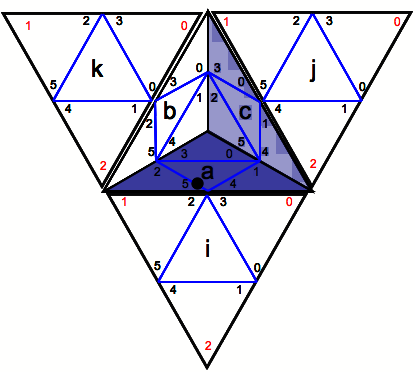}
\caption{When a tetrahedron undergoes deletion, the connectivity must be redefined.  Here, the tetrahedron is being replaced by triangle {\bf a}.\label{deletionex}}
\end{figure}
\end{center}

When a tetrahedron is deleted, three triangles are replaced with one. The triangle with the lowest index is retained (in Figure~\ref{deletionex}, let this be triangle ${\bf a}$ (right)). This triangle replaces the base of the tetrahedron, and so after deletion it becomes triangle {\bf x} in Figure~\ref{deletionex} (left). Only  the coordinates of the apex of this triangle are updated, since it rotates about its base (in the case of Figure~\ref{deletionex}, the $1-2$ edge). The vertices involved may be defined intrinsically, without reference to their embedding space coordinates. The vertex that must be updated is the intersection of the vertices of all three triangles. The new vertex location is the vertex shared by the two triangles that are not the replacement triangle and which is not the apex. In Figure~\ref{deletionex}, that is the vertex shared by triangles {\bf b} and {\bf c}, but not shared by triangle {\bf a}. The coordinates of ${\bf a}_0$ are replaced by the coordinates of ${\bf b}_0$ or vertex ${\bf c}_0$.

The algorithm that performs deletion of tetrahedra depends on the fact that triangles can only be rotated in the surface, they can not be flipped. It is convenient to define deletion in terms of an involution called inversion. When a vector is inverted, the vector is mapped to the other vector that occupies the same edge.  Referring to Figure~\ref{notation}, the pairs are vectors $(0,1)$, $(2,3)$ and $(4,5)$.  If an inversion is performed on vector $3$, we get vector $2$, and so on. This involution is used, together with propagation along the vectors, to redefine the connectivity during deletion using only the intrinsic geometry information   

For example, Figure~\ref{deletionex} shows a tetrahedron that will undergo deletion and be replaced by triangle ${\bf a}$. The connectivity for vectors ${\bf a}_5$, ${\bf a}_0$, ${\bf a}_4$ and ${\bf a}_3$ (where ${\bf a}_i$ for vector $i$ of triangle ${\bf a}$) must be redefined, as they point to triangles ${\bf b}$ and ${\bf c}$ which will be deleted. After deletion, when the tetrahedron is replaced by triangle ${\bf a}$, a particle occupying ${\bf a}_5$ will propagate to ${\bf k}_0$ if undisturbed by collision. Consider the propagation of a fictitious particle from ${\bf a}_5$ to ${\bf b}_4$. After a second propagation this particle would end up on triangle ${\bf c}$, which is incorrect. Inverting the position of the particle, so that it now occupies ${\bf b}_5$ and allowing the particle to propagate once more takes it to ${\bf k}_0$, which is correct. This was achieved by propagating once, inverting, and propagating again. The full set of relabellings given in terms of propagation and inversion are shown in Table~\ref{relabelling}. Vectors ${\bf a}_4$ and ${\bf a}_5$ are readily identified as the vectors attached to the base of the lowest indexed triangle of the tetrahedron. They are updated to point to ${\bf j}_5$ and ${\bf k}_0$ respectively as shown in Table~\ref{relabelling}. The other two vectors may be updated by a similar sequence of propagation and inversion, but it is more straightforward to note that ${\bf a}_0$ is updated to ${\bf k}_1$ which is the inversion of ${\bf k}_0$, and ${\bf a}_3$ is updated to ${\bf j}_4$ which is the inversion of ${\bf j}_5$. Hence the update of vectors ${\bf a}_0$ and ${\bf a}_3$ is obtained by inverting the images of ${\bf a}_5$ and ${\bf a}_4$, respectively.

\begin{table}[h]
  \centering 
\begin{tabular}{lllllll}
\hline
${\bf a}_0$&$\underbrace{\longrightarrow}_\textrm{P}$&${\bf b}_5$&$\underbrace{\longrightarrow}_\textrm{P}$&${\bf k}_0$&$\underbrace{\longrightarrow}_\textrm{I}$&${\bf k}_1$\\
${\bf a}_3$&$\underbrace{\longrightarrow}_\textrm{P}$&${\bf c}_4$&$\underbrace{\longrightarrow}_\textrm{P}$&${\bf j}_5$&$\underbrace{\longrightarrow}_\textrm{I}$&${\bf j}_4$\\
${\bf a}_4$&$\underbrace{\longrightarrow}_\textrm{P}$&${\bf c}_5$&$\underbrace{\longrightarrow}_\textrm{I}$&${\bf c}_4$&$\underbrace{\longrightarrow}_\textrm{P}$&${\bf j}_5$\\
${\bf a}_5$&$\underbrace{\longrightarrow}_\textrm{P}$&${\bf b}_4$&$\underbrace{\longrightarrow}_\textrm{I}$&${\bf b}_5$&$\underbrace{\longrightarrow}_\textrm{P}$&${\bf k}_0$\\
\hline
\end{tabular}
\caption{The sequence of relabellings that occur when the tetrahedron shown in Figure~\ref{deletionex} is replaced in a three-to-one move by the triangle ${\bf a}$. $P$ indicates propagation and $I$ indicates inversion.}\label{relabelling}
\end{table}

A second useful involution is reflection, which returns the vector with opposite velocity on the triangle.  A reflection on vector $1$, for example, returns vector $4$. Because deletion is allowed when ``spectator'' particles are present on the six edges of the replacement triangle it is necessary to update the particle occupancy of the replacement triangle. It is straightforward, in terms of the involutions inversion and reflection, to identify the vectors whose particle occupancy needs to be translated to the replacement triangle.

\subsection{Preventing degenerate triangulations}

In a combinatorial triangulation each triangle is uniquely defined by a set of $3$ vertexes: it is combinatorially unique. Triangulations which do not satisfy this criterion are degenerate. For example, in a degenerate triangulation two vertices may be connected by more than one edge or triangles may share more than one edge. As explained below, unrestricted application of the Pachner moves can result in degenerate triangulations

One form of degeneracy occurs when two of a triangle's neighbors are the same triangle; in other words, two triangles share two edges, or only two triangles meet at a vertex. Such a  feature resembles a ``flap'' attached to the rest of the triangulation. This type of degeneracy is avoided by preventing geometry moves whose post-image contains a flap. We now consider the effect of the two-to-two, one-to-three and three-to-one moves from the point of view of avoiding degenerate triangulations.

Firstly, a flap may be created by the two-to-two Pachner move. The two to two move increases or decreases the number of triangles at a vertex by one.  If three triangles intersect at a point (the apex of a tetrahedron), this will become a flap if two of the triangles are replaced with one.  The application of the two-to-two move to two of the triangles of a tetrahedron will therefore result in a flap.  Preventing two triangles that are part of the same tetrahedron from undergoing a two-to-two move avoids this.  

Second, a degenerate triangulation may not be produced by the one-to-three move (creation). Provided the initial triangulation is not degenerate any vertex that is not at a boundary is shared by at least three triangles. The one-to-three move increases the number of triangles at each existing vertex by at least one, so this move can only create a flap if there were zero triangles to begin with. Hence it is not possible for creation (the one-to-three move) to result in a degenerate triangulation

\begin{figure}[htp]
 \includegraphics[width = 3.0 in]{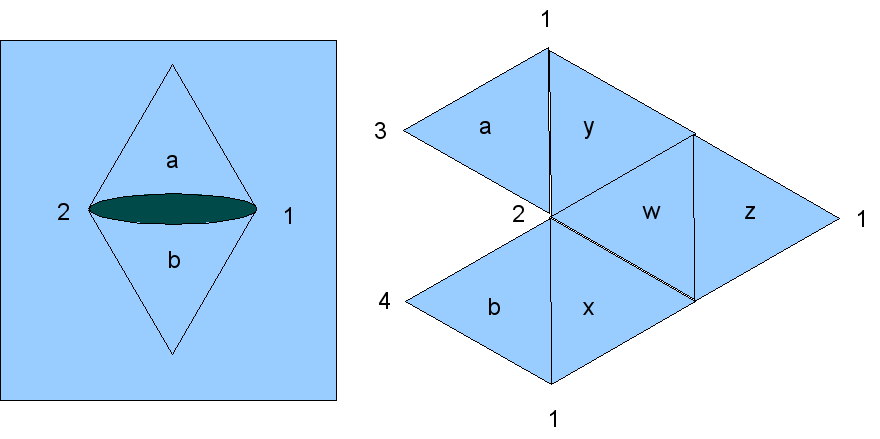}
\caption{A tetrahedron attached by an edge to the main body of the manifold.  If deletion occurs to any triplet of the triangles $w-x-y-z$, a flap will be created.}
\label{flapdemo}
\end{figure}

Third, it is possible for the three-to-one move (deletion) to produce a flap. The three to one move reduces the number of triangles at a vertex by one. This move can create a flap if the geometry as a whole is a tetrahedron, or if a tetrahedron is attached to a manifold by one edge. For example a tetrahedron could be attached to the rest of the manifold via a ``neck''. To make this explicit, take a tetrahedron and label its faces $wxyz$.  Let all the faces be connected except for $x$ and $y$.  Now, take two adjacent triangles, $a$ and $b$, in the manifold that are not connected to each other, and glue the loose edge of $y$ to triangle $a$, and glue the loose edge of $z$ to triangle $b$. A triangulation with a feature like this is degenerate because the two vertices at the join between the tetrahedra and the rest of the triangulation are connected by two edges. If the tetrahedron that is attached by one edge of two triangles were to undergo a deletion of three of its faces, a flap would be created. This is illustrated in Figure~\ref{flapdemo}.

This type of structure can be produced from an initial geometry which is a {\em tetraspiral} - the triangulation which results from successive reflections of each vertex of a tetrahedron through the opposing face~\cite{Tetraspiral1}. In this geometry, and any geometry composed of tetrahedra sharing faces, a single three-to-one move results in two tetrahedra connected by a single edge. A second three-to-one move will then result in a flap.

Such degenerate triangulations are prevented from forming by two checks. First one prevents degeneracy caused by two-to-two moves by checking prior to a two-to-two move that both triangles involved do not belong to the same tetrahedron. This prevents the formation of flaps, and prevents the formation of tetrahedra attached by a shared edge. Second, we check that the three-to-one move is only applied to tetrahedra which are attached to the rest of the triangulation by three edges. It is straightforward to verify that flaps are not produced in our simulations by verifying, for example, that the vertex degree of all triangles is bounded below by three. 

How does the prevention of such degenerate triangulations affect the time reversibility of the model? Consider the inversion of a sequence of moves involving the creation of a degenerate triangulation. Then disallowing the formation of degenerate triangulations corresponds to disallowing degenerate triangulations in the preimage of the inverse rule. If such triangulations are not allowed to form, one must naturally forbid them in the initial geometry.  Disallowing degenerate triangulations in the initial condition is sufficient to maintain reversibility. If such triangulations are allowed in the initial data, but not in the dynamics, invertibility is violated because states exist with two preimages: e. g. one where a flap has been removed due to addition, and one where it was prevented from forming.

\section{Geometry dynamics}

In this section we study some aspects of the dynamics of the geometry degrees of freedom. In these simulations the fluid represented by the particles is quiescent - there is no forcing applied and because the initial velocities of the particles are assigned randomly the average hydrodynamic velocity field will be zero. In one dimension, where the only geometrical degree of freedom is the size of the lattice, both numerical simulation and calculations for particular sets of initial conditions result in an average growth of the lattice size of $t^\frac{1}{2}$~\cite{1D1,1D2,1D3}. We perform the comparable calculations and mean field theory treatment of the two dimensional model. In addition, because the varying vertex degree of the triangulation represents a local geometrical degree of freedom we also present preliminary results on the distribution of vertex degree on the manifold.

\subsection{Mean field theory}

In this section we consider the average behavior of the number of triangles in the model as a function of time. First note that deletion is a rare event compared to addition. Addition requires exactly three particles in one of two configurations of a single triangle. Deletion requires exactly three particles in one of two configurations of three triangles. If all configurations of a triangle occur with equal probability deletion will be less likely than addition simply because it requires correlations between the states of more than one triangle. Since addition is more common than deletion, both by this argument and by observation of actual simulations, we construct a mean field prediction for the behavior of our system with only addition of geometry.  Mean field predictions tend to fail for low dimensional systems.  In the one-dimensional case, for example, the lattice grew as $t^\frac{1}{2}$, but a mean field model predicted $t^\frac{1}{3}$~\cite{1D2}. It is therefore of interest to determine the validity of the mean field prediction in two dimensions.

Let $N$ represent the number of particles and $S$ represent the number of triangles.
The mean number of particles per site is given by
\begin{equation}
\rho = \frac{N}{S}~~~~0\leq \rho\leq 6.
\end{equation}
The probability for a site to undergo addition, $P_+$, will be proportional to the probability that three sites are occupied and three sites are unoccupied.  
\begin{equation}
P_+\propto \rho^3(1-\rho)^3.
\end{equation}
The expected number of triangles which will undergo addition is
\begin{eqnarray}
\langle S_{+}\rangle &=& SP_+ \propto S\rho^3(1-\rho)^3.
\end{eqnarray}
The expected change in the number of triangles is given by
\begin{eqnarray}
\Delta S&=&2\langle S_{+}\rangle\propto S\rho^3(1-\rho)^3.
\end{eqnarray}
\begin{eqnarray}
\Delta S &\propto&  \left(\frac{N^3}{S^2}-3\frac{N^4}{S^3} + 3\frac{N^5}{S^4} + \frac{N^6}{S^5}\right)
\end{eqnarray}
In the limit in which creation dominates deletion and the number of particles, $N$, is conserved, the first term in the equation above will dominate.  Disregarding the last three terms, which will become small as the number of triangles, $S$, grows, we convert this to a differential equation and solve:
\begin{eqnarray}
\frac{dS}{dt} &\propto& \frac{N^3}{S^2}\\
S &\propto& t^{\frac{1}{3}} 
\end{eqnarray}
The mean field prediction is therefore that the lattice will grow asymptotically as $t^\frac{1}{3}$.

\subsection{Results}

Four different types of simulations were performed, each without restrictions upon the curvature of the manifold (except those arising from forbidding degenerate triangulations) or the embedding. All simulations began from an initial icosahedral geometry in which each triangular face was subdivided into $16$ triangles by repeatedly bisecting the edges. Simulations were performed with only creation, only creation and deletion, only creation and the two-to-two rule, and a simulation with creation, deletion, and the two-to-two rule. While simulations with creation but not deletion are not time reversible, they test the hypothesis that the geometry dynamics is dominated by addition. Thirty realizations were performed for each type of simulation with $10^5$ timesteps each to determine the number of triangles as a function of time. The data was fitted to a power law, $S(t) = at^b$, where $S(t)$ represents the number of triangles, in the lattice and $t$ is the number of timesteps.  Fitting the data to the form $S(t) = a(t-t_0)^b$ gave values of $t_0$ of order one, showing that there is only a short transient before the power law growth begins, and so fitting the data to $S(t) = at^b$ is appropriate.

To evaluate the goodness of the fit $\chi^2$ per degree of freedom for the fit function $L(t) = at^b$ was computed:
\begin{equation} 
 \chi^2(a,b) = \frac{1}{n-p}\sum_i \frac{\left[\langle S(t_i)\rangle-f(t_i,a,b)\right]^2}{\sigma_i^2}
\end{equation}
where $n$ is the number of data points, $p$ is the number of parameters, in this case 2, $a$ and $b$, and $\sigma_i$ is the standard deviation on the mean for each $\langle S_i\rangle$, where
\begin{equation}
 \langle S_i\rangle = \frac{1}{r}\sum_j S_{ij}.
\end{equation}
and $j$ is the number of realizations.

\begin{figure}[htp]
\centering
\subfigure[]{\includegraphics[width=2.8in]{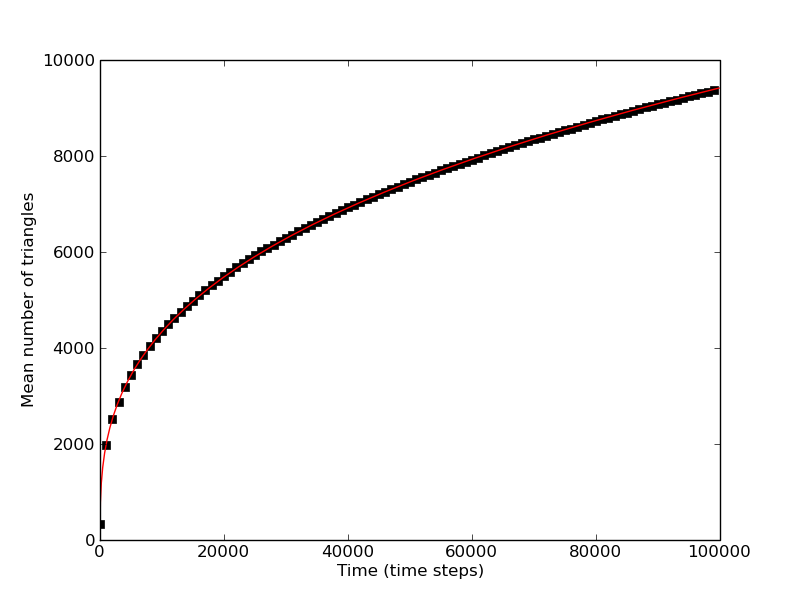}}
\subfigure[]{\includegraphics[width = 3.0 in]{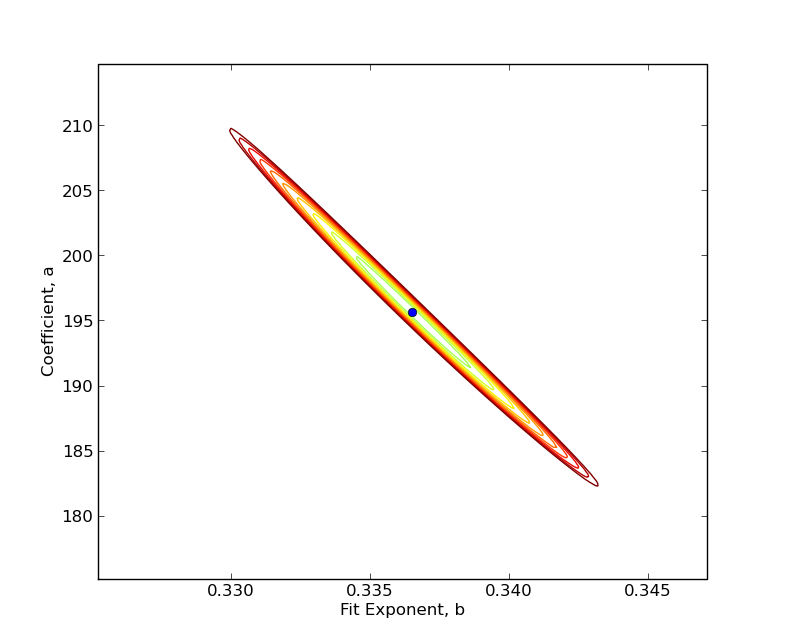}}
\caption{{\bf All Pachner moves allowed.}  Number of triangles as a function of time averaged over $30$ realizations for $100000$ timesteps each with no restrictions on the curvature or embeddability of the triangulation, and all Pachner moves utilized. The symbols in a) show the simulation data every $1000$ timesteps and are larger than one standard deviation of the mean. The solid line is a fit created in Origin 7.0 with a Levenberg-Marquardt method for $L(t) = at^b$ to the complete data set of $100000$ points. Figure b) shows a contour plot of $\chi^2$ using a sampling of $200$ points evenly spaced along the range of $a$ and $b$. The minimum $\chi^2$ value lies at $\chi^2 = 0.0319$ at $a=196$ and $b=0.33618$ in agreement with the fit found by Origin 7.0 and the outer-most contour represents a deviation of $1.0$ from this minimum.  The fitted value of the exponent is $b=0.33618\pm 0.0065$, consistent with a power law exponent of $1/3$.\label{allmoves}}  
\end{figure}

\begin{figure}[htp]
\subfigure[]{\includegraphics[width=3.0in]{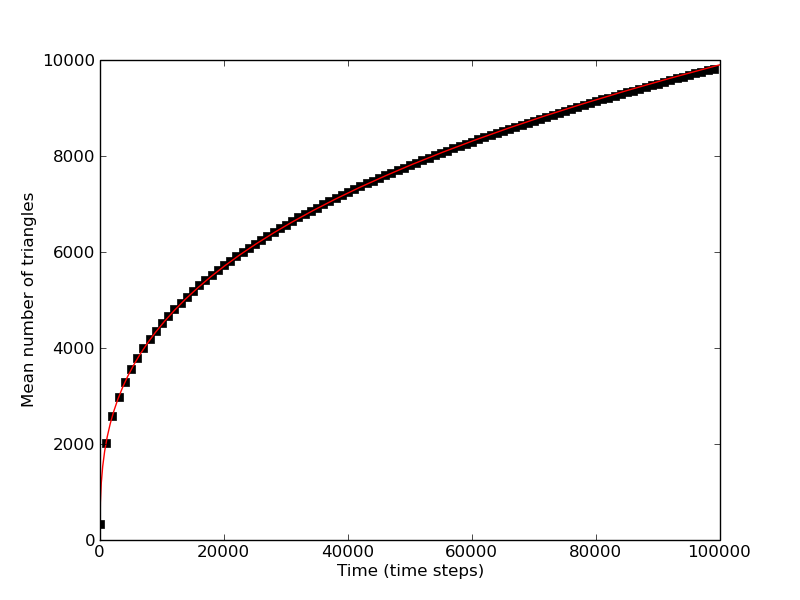}}
\subfigure[]{\includegraphics[width =3.0 in]{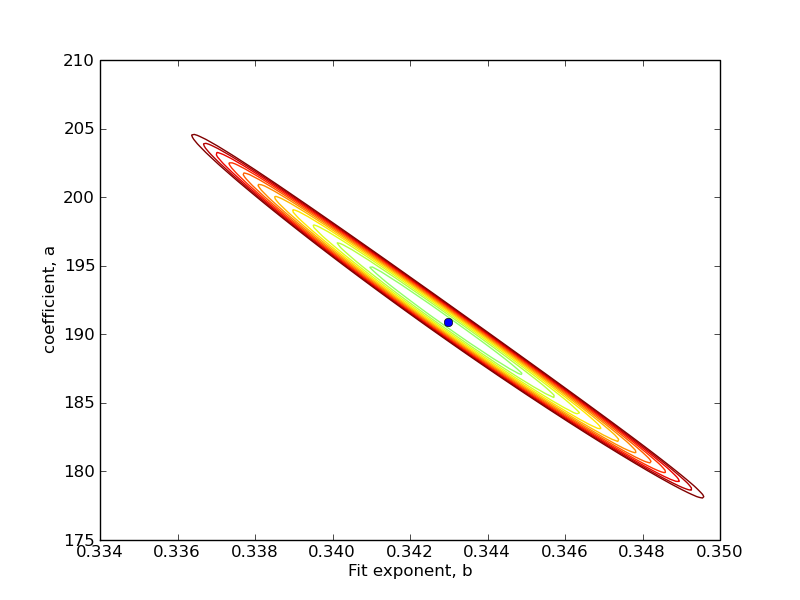}}
\caption{{\bf Only addition of tetrahedra.} Number of triangles as a function of time averaged over $30$ realizations for $100000$ timesteps each. The simulations included only the one-to-three Pachner move with no growth or embedding control. The symbols in a) show the simulation data every $1000$ timesteps and are larger than one standard deviation of the mean. The solid line is a fit created in Origin 7.0 with a Levenberg-Marquardt method for $L(t) = at^b$ to the complete data set of $100000$ points. Figure b) shows a contour plot of $\chi^2$ using a sampling of $200$ points evenly spaced along the range of $a$ and $b$. The minimum $\chi^2$ value lies at $\chi^2 = 1.106$ at $a=190.864$ and $b=0.343$ in agreement with the fit found by Origin 7.0 and the outer-most contour represents a deviation of $1.0$ from this minimum.  The fitted value of the exponent is $0.337\leq b=0.344\leq 0.350$, consistent with a power law exponent of $1/3$.\label{onlyadd}}  
\end{figure}

\begin{figure}[htp]
\includegraphics[width=3.0in]{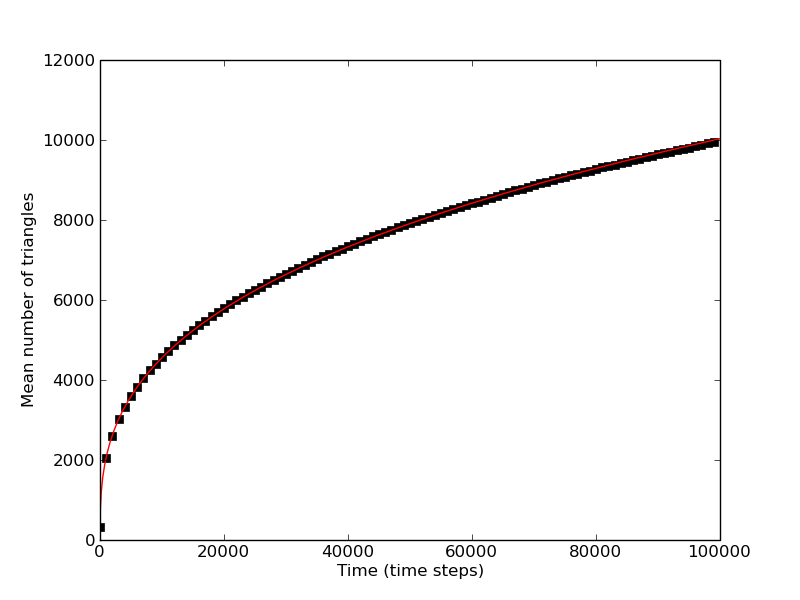}
\includegraphics[width =3.0 in]{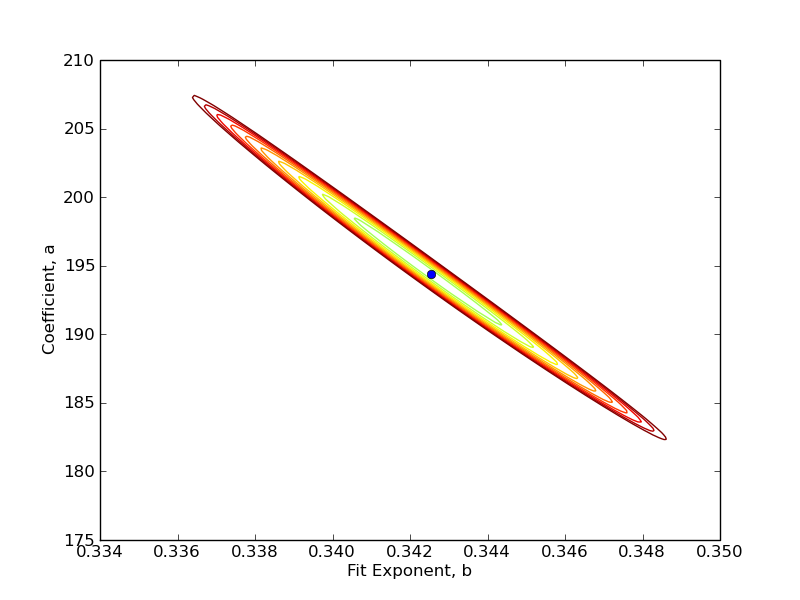}
\caption{{\bf No deletion.} Number of triangles as a function of time averaged over $30$ realizations for $100000$ timesteps each. The simulations included both two-to-two and one-to-three Pachner moves with no growth or embedding control. The symbols in a) show the simulation data every $1000$ timesteps and are larger than one standard deviation of the mean. The solid line is a fit of $L(t) = at^b$ to the complete data set of $100000$ points. The fit was created in Origin 7.0 using a Levenberg-Marquardt method. Figure b) shows a contour plot of $\chi^2$ using a sampling of $200$ points evenly spaced along the range of $a$ and $b$. The minimum $\chi^2$ value lies at $\chi^2 = 0.878$ at $a=194.402$ and $b=0.343$ in agreement with the fit found by Origin 7.0. The outer-most contour represents a deviation of $1.0$ from this minimum.  The fitted value of the exponent is $0.336\leq b=0.343\leq 0.349$, consistent with a power law exponent of $1/3$.\label{nodel}.}
\end{figure}

\begin{figure}[htp]
\subfigure[]{\includegraphics[width=3.0 in]{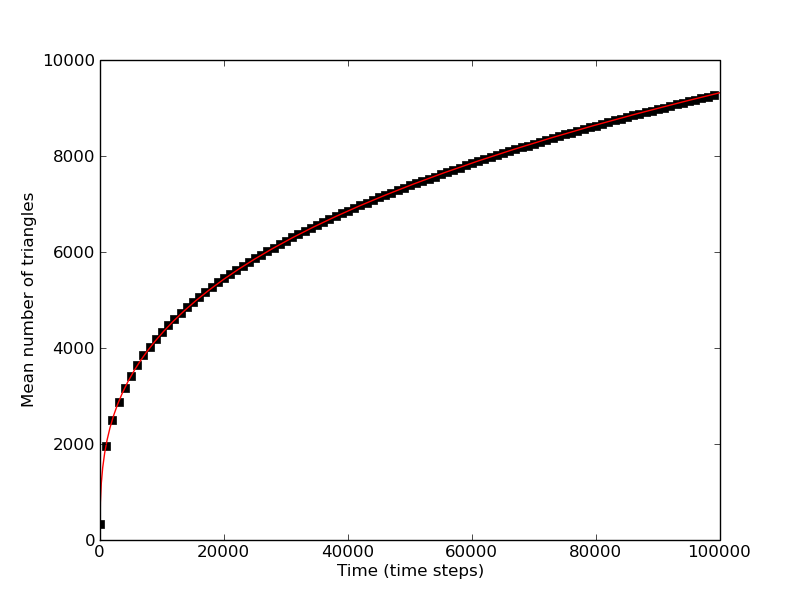}}
\subfigure[]{\includegraphics[width =3.0 in]{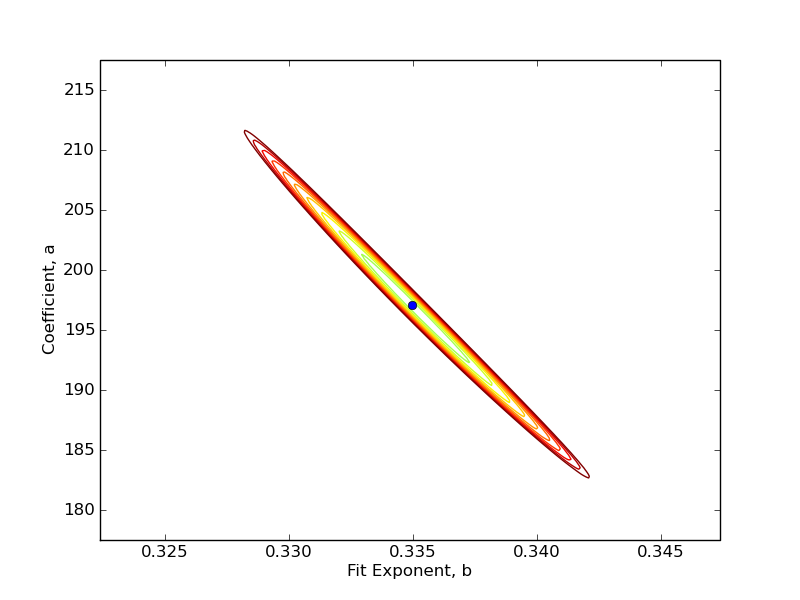}}
\caption{{\bf No two-to-two moves allowed.} Number of triangles as a function of time averaged over $30$ realizations for $100000$ timesteps each. The simulations included both the one-to-three and three-to-one Pachner moves with no two-to-two moves and without restrictions on the curvature or embeddability of the triangulation. The symbols in a) show the simulation data every $1000$ timesteps and are larger than one standard deviation of the mean. The solid line is a fit of $L(t) = at^b$ to the complete data set of $100000$ points. The fit was created in Origin 7.0 using a Levenberg-Marquardt method. Figure b) shows a contour plot of $\chi^2$ using a sampling of $200$ points evenly spaced along the range of $a$ and $b$. The minimum $\chi^2$ value lies at $\chi^2 = 0.666$ at $a=197.060$ and $b=0.335$ in agreement with the fit found by Origin 7.0. The outer-most contour represents a deviation of $1.0$ from this minimum.  The fitted value of the exponent is $0.330\leq b=0.335\leq 0.342$, consistent with a power law exponent of $1/3$.\label{no22}}  
\end{figure}

For each type of simulation an initial fit using Origin 7.0~\cite{origin7} was obtained (using a Levenberg-Marquardt method) and an independent error analysis was performed by computing $\chi^2$ in the $a, b$ plane. The minimum value of $\chi^2$ found via this method matches that found by Origin 7.0. The parameter uncertainties were obtained by this $\chi^2$ analysis by allowing $\chi^2$ to increase by one above the minimum. The uncertainties so obtained are larger than those given by Origin, presumably because we allow $a$ and $b$ to vary independently. In all four types of simulation the exponent value is consistent with a power law exponent of $1/3$, in agreement with the mean field prediction. The data and $\chi^2$ analysis for the simulations with all Pachner moves is shown in Figure~\ref{allmoves}, that for the simulations with only the three-to-one (addition) Pachner move is shown in Figure~\ref{onlyadd}, with all Pachner moves except the three-to-one move (deletion) in Figure~\ref{nodel} and with all Pachner moves except the two-to-two move in Figure~\ref{no22}

\subsection{Curvature analysis}

\begin{figure}[htp]
\includegraphics[width=3.0 in]{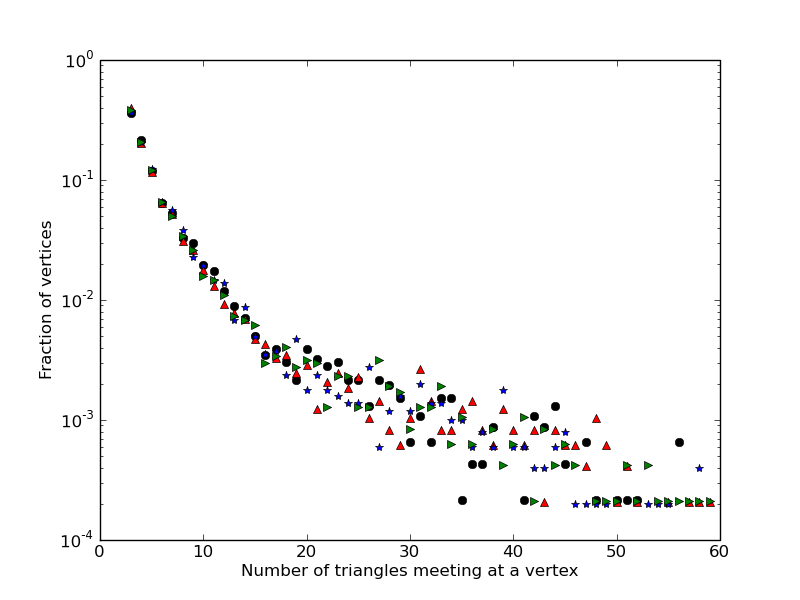}
\caption{Histogram of number of triangles meeting at each vertex. Results of four separate simulations for $100000$ timesteps. The initial geometry was an icosahedron with each of its faces subdivided into $16$ equilateral triangles. Data is shown for one simulation with all Pachner moves implemented (Black circles), one simulation with only the three-to-one addition move implemented (Red upward pointing triangles), one simulation with the one-to-three addition move and the two-to-two move but no deletion (Blue stars) and one simulation with addition and deletion moves but no two-to-two move (Green sideways pointing triangles).\label{curvehist}} 
\end{figure}

Unlike the one-dimensional model it is possible to define a curvature variable at each vertex of the triangulation. As we do not restrict our triangulations in any way in the simulations described above it is of interest to quantify how curvature is distributed for a typical realization. We performed four simulations of a single realization of the type of simulation displayed in Figures~\ref{allmoves}, \ref{onlyadd},\ref{nodel}, \ref{no22}. A histogram of the vertex degree is shown in Figure~\ref{curvehist}. As expected, by allowing unrestricted addition of tetrahedran vertices of arbitrarily high degree form in the triangulation. However, most of the vertices of the triangulation have degree between three and ten. While the data shown in Figure~\ref{curvehist} is insufficient to support a detailed quantitative analysis of the distribution of vertex degree, it appears by inspection to be consistent with an exponential distribution for degrees between three and ten. Larger vertex degrees appear to be more common than that predicted by this trend below vertex degree $10$, but there is insufficient data to draw conclusions here.  If we denote the number of tetrahedra added to the original geometry $N_1$, and tetrahedra added to these $N_2$ and so on, an exponential distribution is consistent with the ratio of $N_i$ to $N_{i+1}$ being a constant.

\begin{figure}[htp]
\subfigure[]{\includegraphics[width=3.0 in]{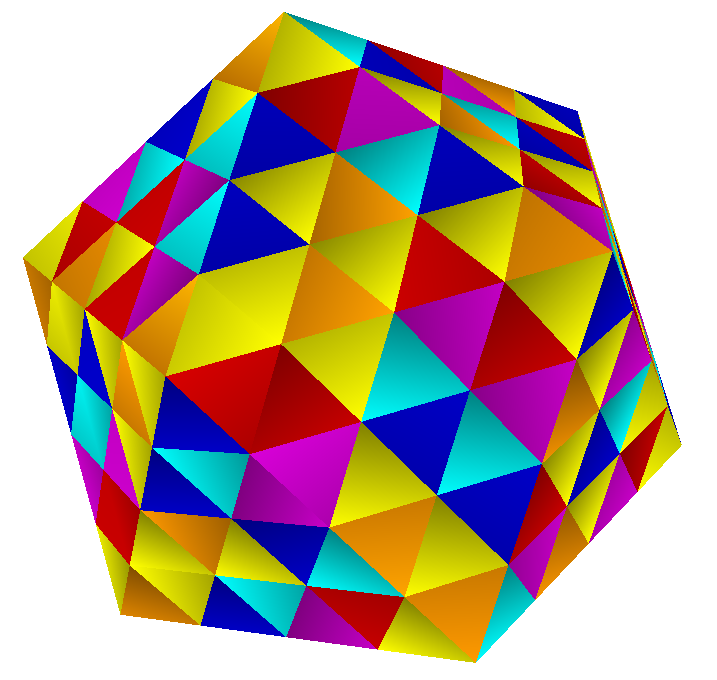}}
\subfigure[]{\includegraphics[width=3.0 in]{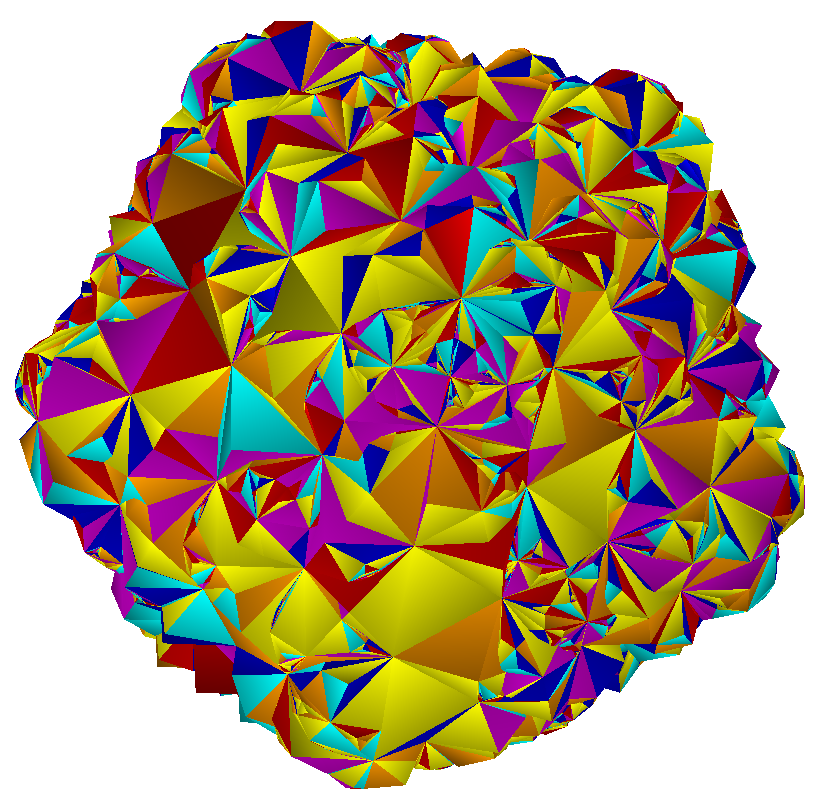}}
\caption{Visualizations of the triangulation for a typical realization. a) Initial condition for all simulations. b) Typical geometry after $100000$ time steps for a simulation with only the three-to-one Pachner move (addition) implemented, with no restriction on the curvature. The faces shown here are individual triangles - with only the three-to-one Pachner move implemented the embedding is always isometric. The appearance of many smaller triangles is due to the fact that the surface is extensively self-intersected.\label{visual}}  
\end{figure}

In Figure~\ref{visual} we display visualizations of the triangulation for a typical realization. This figure shows a simulation in which only the three-to-one Pachner move is implemented, resulting in a triangulation which is always isometrically immersible in three dimensions. This simulation shows that these triangulations self intersect many times. This results in the appearance of many small triangles in the visualization - in fact these are parts of triangles which intersect each other. Visualizations of realizations with any combination of rules applied all showed this feature. In addition, one observes clusters of added tetrahedra in all simulations. This may be due to the fact that added tetrahedra act as scattering centers for particles, and so groupings of tetrahedra will naturally increase the probability of further geometry-changing collisions by causing particles to spend longer in a given region.

\section{Conclusion}

We have presented the first lattice-gas model with dynamical geometry in two dimensions. Our model is an extension of the FHP hydrodynamic two-dimensional lattice gas model, and the one-dimensional dynamical geometry lattice gas~\cite{FHP,1D1,1D2,1D3}. We have defined and implemented rules for dynamical geometry by both Pachner moves. For a quiescent fluid on an initially icosahedral geometry the number of triangles grows as $t^{1/3}$ for all combinations of rules simulated. This is in agreement with a mean field prediction, a fact of some interest as mean field predictions generally fail in low dimensions and in fact fail for the one dimensional version of this model~\cite{1D2,1D3}.

Unlike the one-dimensional case, the flat space limit of this model is non-trivial: it is the hydrodynamic FHP lattice gas. For our model as defined it is therefore possible to perform simulations in three regimes. Firstly, the fluid may be quiescent, and the geometry dynamical, the limit studied in this paper. Secondly, the geometry may be fixed and non-trivial, and the fluid driven. This regime is relevant for the simulation of surface flows on fixed background geometry, such as atmospheric flows on the sphere, and in experiments with curved soap films~\cite{seychelles:144501}. Thirdly, the fluid may be driven and the geometry dynamical, a situation of relevance to surface flows in fluid interfaces. Indeed, the equations for surface flow are well known, including the case in which the  surface is dynamical~\cite{Scriven,greenbook}.

Given the model defined here, a natural question to pose is: what are the macrodynamical equations of motion? In the regime where a non-trivial flow occurs on a fixed background geometry, are the relevant fluid equations the Navier-Stokes equations on the manifold represented by the triangulation? In the regime where the geometry is dynamical, does the time evolution of the flow coupled to the geometry obey the continuum equations of surface flow given in~\cite{Scriven,greenbook}?

The principal tool used to obtain the macrodynamical equations of a given lattice-gas model is the Chapman Enskog expansion~\cite{chapman}. This is an asymptotic expansion around a local equilibrium distribution. It is valid in the regime that local equilibrium, characterized by a few hydrodynamic fields, is reached rapidly, while global equilibration occurs on longer timescales by hydrodynamic processes. Analysis of the model defined in this paper requires a new variation of the Chapman Enskog  expansion. 

To treat the model on a fixed, curved surface the Chapman Enskog analysis would need to be extended to arbitrary two-dimensional manifolds. To treat the case where the geometry become dynamical it must be possible to introduce the geometry degrees of freedom into the Chapman Enskog analysis. One way to do this is to define the continuum limit of the triangulation in the same way as the continuum limit of the velocity field. That is, one considers an average (time, spatial or ensemble) over many triangulations of the same surface. The macrodynamical equations of surface flow given in~\cite{Scriven,greenbook} might then arise, for suitably chosen collision rules, from a Chapman-Enskog analysis as the slow relaxation of fluid plus geometry after a fast relaxation to an equilibrium geometry. If such an analysis is valid for the model defined in this paper, it would also allow simulation of fixed geometries via simulation of an ensemble of dynamical geometries fluctuating about an average continuum surface. 

The equilibrium statistical mechanics of two dimensional triangulated surfaces embedded in three dimensions has been well studied~\cite{KKN1,KKN2,Wheater}. The model defined here differs from this body of work in several ways. The tethered surfaces studied in~\cite{KKN1} have a fixed internal metric and a Hamiltonian which depends only on extrinsic embedding coordinates. The triangulations of our model have an intrinsic metric which varies dynamically due to application of the Pachner moves. In the terminology of~\cite{KKN1} this makes our surfaces liquid rather than tethered. The object of study of~\cite{KKN1} and subsequent work was the equilibrium properties of embedded surfaces, here we are interested in the non-equilibrium dynamics of surfaces on which there is a non-trivial vector field whose dynamics is coupled to the intrinsic geometry of the triangulation.

However, for the case simulated in this paper in which the fluid degrees of freedom are quiescent it is interesting to compare the typical geometries shown in Figures~\ref{visual} with the equilibrium geometries in the crumpled phase of random surface models. Two observations are relevant. Firstly, as we allow self intersection and do not restrict the embedding our surfaces are phantom surfaces and should be compared with random surface models which have no extrinsic curvature term in their Hamiltonian. The equilibrium geometries of such random surface models are crumpled and contain many ``spiky'' features. 

The typical geometries in our simulations exhibit similar features - the high degree vertices shown in Figure~\ref{curvehist} occur at branching points where many tetrahedra share a common vertex. The geometries shown in Figure~\ref{visual} also exhibit a concentration of new tetrahedra - showing that tetrahedra are added on new tetrahedra more often than on the original geometry. This can be explained by the fact that the curvature produced by new geometry will act as scattering centers for the particles - causing particles to spend longer in the vicinity of new geometry, where they will then scatter and add further new gemetry. This will naturally lead to a branched polymer-like structure where tetrahedra are added to tetrahedra and particles become trapped on the new branches of the geometry.

Future study of the model should determine whether the geometries produced by the model without constraint are indeed in the crumpled phase. Simulations in which constraints are applied to the local curvature or embedding of the triangulation may result in geometries closer to smooth manifolds and so may be necessary for applications in which one aims to simulate a fluid moving on a smooth two dimensional surface.

Fluids, while frequently treated as continua, are in fact composed of atoms or molecules. The lattice-gas and lattice-Boltzmann methods use the existence of an underlying statistical description of a fluid to realize efficient numerical methods for fluid simulation~\cite{redbook,succibook}. While a discretization of space and time underpins most numerical methods for field theories, our most fundamental current understanding is that spacetime is a continuous Lorentzian manifold. 

The idea that, like fluids, spacetime may have underlying discrete substructure occurs repeatedly in speculative models for quantum gravity. A treatment of classical general relativity on polyhedral simplicial complexes was first considered by Regge~\cite{Regge}. In the causal dynamical triangulations approach the four space-time dimensions emerge from an ensemble of simplicial complexes, suitably constrained by causality~\cite{Ambjorn2}. In the causal sets approach the underlying Lorentzian manifold is replaced by a discrete set of points with a causal (partial) order~\cite{Bombelli}. In loop quantum gravity geometrical operators such as area and length have a discrete spectrum~\cite{rovellibook}. In discrete models of quantum gravity the apparently continuous classical spacetime emerges at large scales due to the smallness of the Planck length. In the more experimentally accessible world of fluid dynamics, the continuum picture is valid because of the largeness of Avagadro's number. The discrete model of fluid mechanics on arbitrary triangulated surfaces presented here provides a model in which the question of the emergence of smooth manifold-like structures, and associated dynamics of a classical field on the manifold, may be studied without the numerous conceptual problems of both general relativity and quantum mechanics.

\section*{Acknowledgements}

This work received financial support from Research Corporation for Science Advancement through a Cottrell College Science award, and from the Sherman Fairchild Foundation and Howard Hughes Medical Institute. PJL thanks Bruce Boghosian, Gianluca Caterina, Suzanne Amador-Kane and Stephon Alexander for stimulating discussions, and the Department of Mathematics at UCSD and the Institute for Quantum Information at Caltech for hosting visits during which parts of this work were completed.


\end{document}